\documentstyle[12pt,aaspp4,flushrt]{article}

\newcommand\psr{PSR~J1105$-$6107}
\newcommand\snr{G290.1$-$0.8}
\newcommand\etal{et al.~}
\newcommand\gro{2EG J1103$-$6106}

\begin{document}

\title{Discovery of the Young, Energetic Radio Pulsar PSR J1105$-$6107}

\author{V. M. Kaspi\altaffilmark{1}}
\affil{Massachusetts Institute of Technology, Physics Department, Center for Space Research 37-621, 70 Vassar Street, Cambridge, MA, 02139}
\authoremail{vicky@space.mit.edu}

\author{M. Bailes}
\affil{University of Melbourne, School of Physics, Parkville, Victoria 3052, Australia}
\authoremail{mbailes@isis.ph.unimelb.edu.au}

\author{R. N. Manchester}
\affil{Australia Telescope National Facility, CSIRO, PO Box 76, Epping NSW 2121, Australia}
\authoremail{rmanches@atnf.csiro.au}

\author{B. W. Stappers}
\affil{Mount Stromlo and Siding Spring Observatories, ANU, Private Bag, Weston Creek, ACT 2611, Australia}
\authoremail{bws@merlin.anu.edu.au}

\author{J. S. Sandhu}
\affil{California Institute of Technology, MS 105-24, Pasadena, CA 91125}
\authoremail{jss@astro.caltech.edu}

\author{J. Navarro\altaffilmark{2}}
\affil{National Radio Astronomy Observatory, P.O. Box 0, Socorro, NM 87801}
\authoremail{j.navarro@ieee.org}

\author{N. D'Amico}
\affil{Osservatorio Astronomico, via Zamboni 33, 40126 Bologna and
Istituto di Radioastronomia, Via P. Gobetti 101, 40129 Bologna, Italy}
\authoremail{damico@astbol.bo.cnr.it}

\altaffiltext{1}{Hubble Fellow}
\altaffiltext{2}{Current address:  Kvednaberget 12, 4033 Forus, Norway}

\begin{abstract}

We report the discovery and follow-up timing observations of the
63 ms radio pulsar, \psr.  The pulsar is young, having a characteristic
age of only 63~kyr and, from its dispersion measure, is estimated to be 
at a distance of $\sim$7~kpc from the Sun. We consider its possible association 
with the nearby supernova remnant \snr\ (MSH 11$-$6{\it 1}A);
an association requires that the pulsar's proper motion be $\sim$22
mas~yr$^{-1}$ (corresponding to $\sim$650 km~s$^{-1}$ for a distance of
7~kpc) directed away from the remnant center, assuming that
the characteristic age is
the true age.  The pulsar's spin-down luminosity, $2.5 \times
10^{36}$~erg~s$^{-1}$, is in the top 1\% of all known pulsar spin-down
luminosities.  Given its estimated distance, \psr\ is therefore likely to
be observable at high energies.  Indeed, it is coincident with the
known CGRO/EGRET source \gro; we consider the possible association and
conclude that it is likely.

\end{abstract}


\section{Introduction}

We have discovered a 63~ms radio pulsar, PSR~J1105$-$6107, during a 
recent search for
pulsars using the 64~m radio telescope at Parkes, NSW, Australia.  Most
pulsars with similar rotation periods fall into one of two categories:
those that are young and energetic, the short spin period a result of
a relatively recent birth,
or those that have been mildly recycled by a binary
companion, like the original binary pulsar, PSR B1913+16, which has a spin
period of 59~ms. Thus, a 63~ms radio pulsar is an important find, worthy
of further investigation.

Here we report on the discovery and follow-up timing observations of
\psr, and show that it is a member of
the first category above, namely young and energetic.  Pulsars are
hypothesized to have been born in supernovae; the existence
of a young pulsar therefore requires us to consider whether there is an
associated remnant of a supernova explosion.  In the case of
\psr, we consider its possible association with the known remnant \snr\ 
(MSH 11$-$6{\it 1}A) which lies nearby on the sky.
In addition, the existence of
an energetic pulsar requires us to consider whether there should be
associated high-energy emission, since magnetospheric X-ray and
$\gamma$-ray emission, generally representing a significant fraction of
the pulsar's spin-down luminosity, very likely holds clues to the
yet-elusive pulsar emission mechanism.  We show that \psr\ should be
detectable at high energies, and consider whether the known
EGRET $\gamma$-ray source \gro\ is associated with the pulsar.

\section{Observations and Results}

\psr\ was discovered with the Parkes telescope in 1994 July in a search at 
a central radio 
frequency of 1420~MHz.  The search targeted OB runaway stars in the hope
of detecting new pulsar/OB star binaries.  \psr\ was discovered while
pointing at the B4V star HD96264, but the follow-up observations described 
below rule out any association.  The details
of the search will be described elsewhere (\cite{kmd97}).

\begin{figure}
\plotone{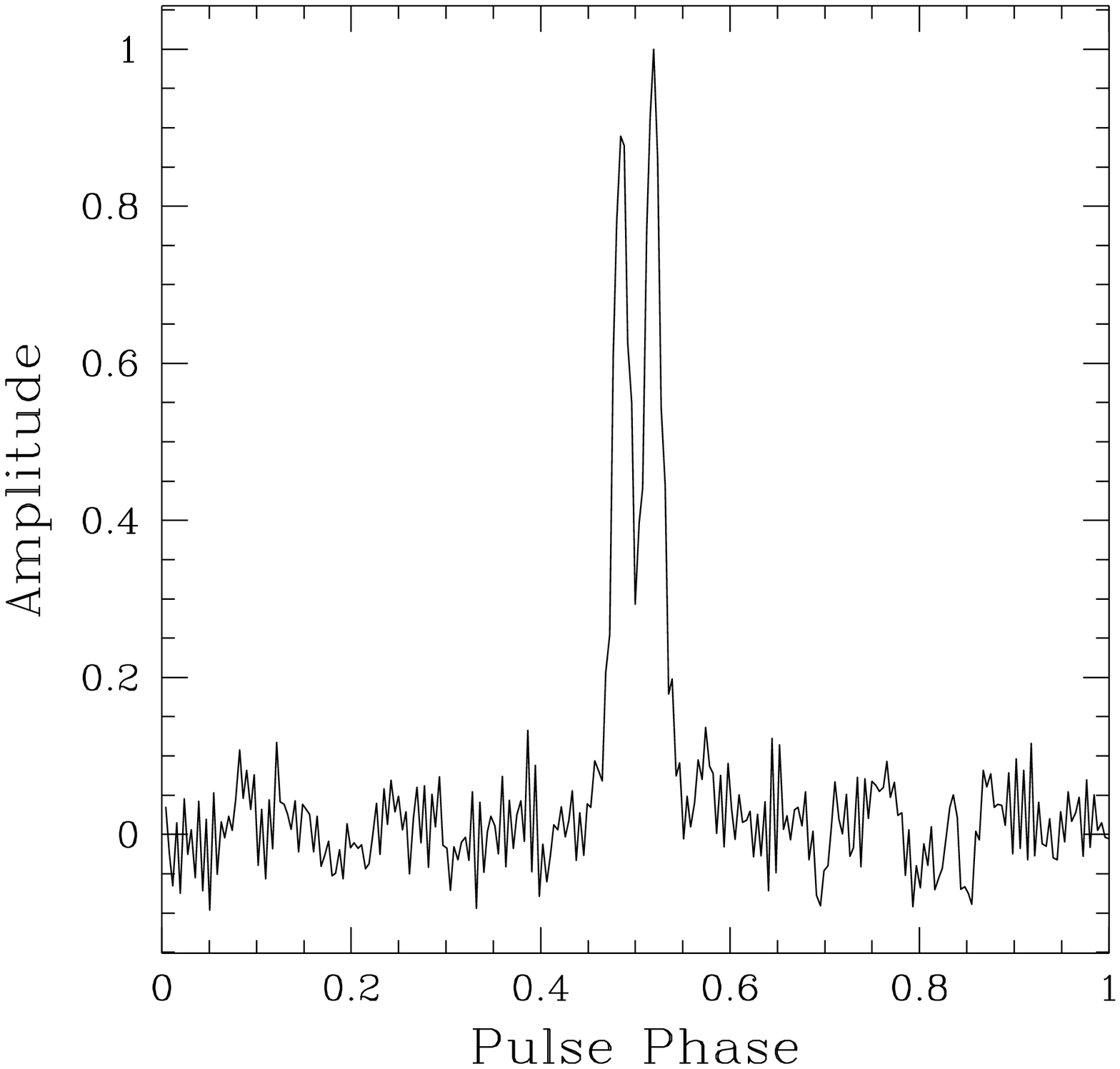}
\caption{Pulse profile for \psr\ at 1650~MHz, with peak normalized to
unity.  The profile is similar at
1420~MHz.  The apparent off-pulse structure is a result of interference.}
\label{fig:prof}
\end{figure}

A total of 96 timing observations of \psr\ were
obtained between 1993 July 8 and 1996 July 4 at Parkes.  
Of these, 91 were obtained
at central radio frequencies ranging from 1390 to 2050~MHz.  
The remaining five observations were obtained at 660~MHz in 1995 July.
All data taken before 1995 were obtained
using filter-bank timing systems (2 $\times$ 64 $\times$ 5~MHz at 1520~MHz
and 2 $\times$ 256 $\times$ 0.125~MHz at 660~MHz) that have been described in
detail elsewhere (e.g. \cite{bhl+94}).  Most data
from after 1995 were obtained using the Caltech correlator-based pulsar timing 
machine (\cite{nav94}), which has $2 \times 128$ lags across 128~MHz in each of
two separate frequency bands.  Typically, correlator observations were
made at central frequencies of 1420 and 1650~MHz simultaneously.
Compared to the filter-bank system,
the correlator's narrower frequency channels
resulted in less channel dispersion 
smearing and hence finer time resolution.  Filter-bank data were recorded
on tape and folded off-line; correlator data were folded on-line.
Average profiles were convolved
with high signal-to-noise ratio templates to yield pulse arrival times. 
The average profile at 1650~MHz shown
in Figure~\ref{fig:prof} was obtained by aligning and summing numerous 
individual correlator profiles.  The profile at 1420~MHz is similar, while
the two components obvious in Figure~\ref{fig:prof} cannot be resolved in the
660 MHz data because of dispersion smearing.  
Resulting arrival times were analyzed using
the standard {\sc TEMPO} pulsar timing software package (\cite{tw89})
together with the JPL~DE200 ephemeris (\cite{sta82}).  Typical arrival
time uncertainties were $\sim$250~$\mu$s for $\sim$10~min integrations
with signal-to-noise ratio $\sim$20 at frequencies above 1390~MHz.  Arrival
times at 660~MHz had uncertainties approximately twice as large.

In the timing analysis, the pulsar dispersion measure (DM) was first
determined from delays across the observed bands and then refined
using 15 arrival times measured in 1995 July, including
all five measured at 660~MHz.  This was to ensure  good
frequency coverage, and no contamination from long-term timing
noise so common to young pulsars.  The measured DM is 
($271.01 \pm 0.02$)~pc~cm$^{-3}$
and during the subsequent timing analysis it was
held fixed at this value.  To minimize contamination
of the timing position from long-term timing noise, we ``pre-whitened''
the data (e.g. \cite{ktr94}), 
fitting for sufficiently many frequency derivatives (four) to render
the residuals approximately Gaussian distributed, determined by eye.
The timing position, determined while fitting for these derivatives,
is given in Table~\ref{ta:parms}, and was subsequently held fixed.
Finally, we measured the best
period and period derivative, also given in Table~\ref{ta:parms}.
The uncertainties in all parameters are 
1$\sigma$ statistical uncertainties, obtained assuming equal weighting for all
arrival times.
The pulsar's surface magnetic field
$B = 3.2 \times 10^{19} \; {\rm G} \; (P \dot{P})^{1/2} \simeq 1 \times 10^{12}$~G,
and its spin-down luminosity 
$\dot{E} = 4 \pi^2 I \dot{P} / P^3 \simeq 2.5 \times 10^{36}$~erg~s$^{-1}$, where
the neutron star moment of inertia $I$ is taken to be $10^{45}$~g~cm$^2$.
The pulsar's characteristic age $\tau_c \equiv P/2\dot{P} \simeq 63$~kyr.

\begin{figure}
\plotone{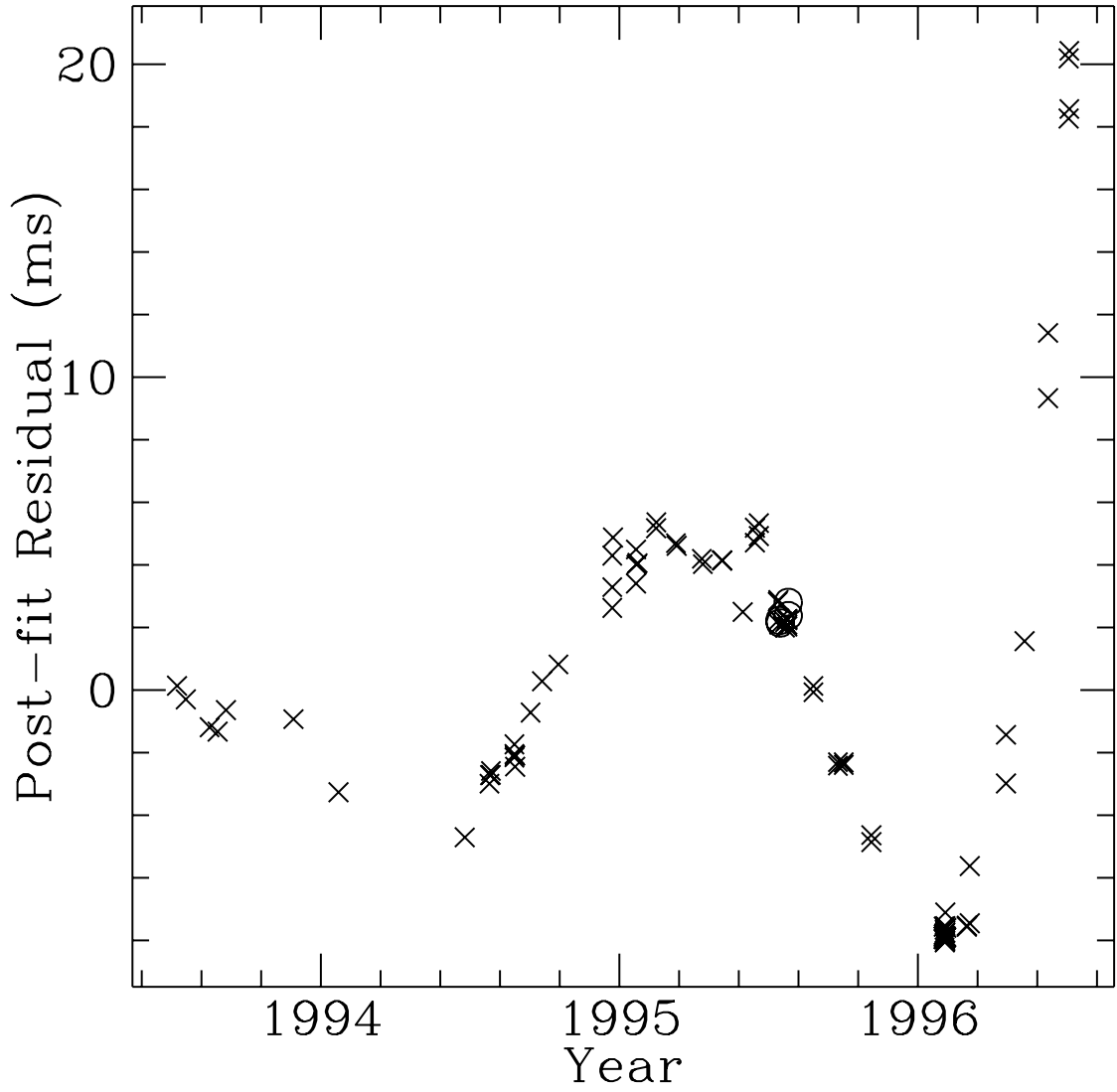}
\caption{Post-fit residuals for \psr\ after subtracting the model
given in Table~\protect\ref{ta:parms}.  Residuals at 660~MHz are shown as open circles.
Uncertainties on arrival times are typically smaller than the size of the symbols.}
\label{fig:res}
\end{figure}

Post-fit residuals, obtained after removing the timing model given
in Table~\ref{ta:parms}, are shown in Figure~\ref{fig:res}.  In the plot,
uncertainties are typically much smaller than the size of the symbol.
The timing noise, interpreted as irregularities in the neutron star's rotation,
is obvious.
We can quantify the amount of timing noise, as
prescribed by Arzoumanian \etal (1994), \nocite{antt94} by measuring 
$\Delta_8 \equiv \log (| \ddot{\nu} | t^3 / 6 \nu)$, 
where $\nu = 1/P$ and for $t = 10^8$~s.
For \psr\ we find $\Delta_8 = -0.8$, which is consistent within the
scatter with $\Delta_8$ parameters for other young pulsars.
The pulsar's small $\tau_c$, as well as the large amount of timing noise,
suggest that \psr\ is an excellent candidate for glitches.  Indeed we cannot
rule out some contamination of the timing parameters by a slowly-relaxing
glitch that occurred before 1993 July (c.f. \cite{mkj+91}; \cite{lkb+96}).

That the timing observations for \psr\ reported here extend back a
full year before the pulsar's discovery requires some
explanation.  In general, raw, dispersed, and unfolded filter-bank
timing data are recorded and archived on tape. 
By chance, \psr\ lies less than one Parkes 1420~MHz primary
beam-width from the pulsar PSR J1103$-$6101, which was discovered  in
1992 July, as part of a major search for pulsars near supernova
remnants (\cite{kmj+96}).  
The DM toward PSR J1103$-$6101 is only 75~pc~cm$^{-3}$,
indicating that it is a foreground object.  Once the discovery of \psr\ was made
and its proximity to PSR J1103$-$6101 realized, the archived raw data
for the latter were retrieved, dedispersed, and folded at the
former's parameters.  
\footnote{\psr\ was not detected as part of the Kaspi \etal (1996)
search because it was outside the Parkes search beam.  Only in 1993 July
was the correct timing position for PSR J1103$-$6101
determined.  Observations made after that date had the telescope
pointing at PSR J1103$-$6101's refined position, which was closer to
that of \psr.}

\section{Discussion}

\subsection{Possible Association with G290.1$-$0.8}

\psr\ is located near the Galactic supernova remnant \snr, also known
as MSH 11$-$6{\it 1}A (\cite{sg70b}).  The proximity of a young pulsar to a supernova
remnant suggests that
they may have been formed in the same explosion.  Alternatively
they may be coincidentally superposed on the sky; the Galactic plane
is replete with pulsars and supernova remnants and the possibility of
chance alignment is non-negligible. 
Indeed a spurious association is not implausible, as the true
remnant of the pulsar's birth may well have faded from view
(\cite{bgl89}), and the explosion that produced \snr\ may not have produced
a neutron star.
Here we consider whether there is a genuine association between \psr\ and
\snr.

The Taylor \& Cordes (1993) \nocite{tc93} DM-distance model places \psr\ at 
a distance of 7~kpc,
given its DM and Galactic coordinates.  The uncertainty on this
distance is estimated to be $\sim$25\%.   Thus, the range
consistent with the DM-distance model is 5--9~kpc.  Regions
of enhanced free-electron density (like H {\sc II} regions) along the
line-of-sight can result in an overestimate of the pulsar's distance
from its DM.

The distance to \snr\ has been estimated many times in the literature.
First, H{\sc I} absorption measurements made by Dickel (1973)
\nocite{dic73} suggest that the
remnant is probably at  3--4~kpc.  The $\Sigma-D$ relation,
which is known to be very uncertain, suggests a distance of 3--6~kpc 
(Clark \& Caswell 1976).
\nocite{cc76}  Other authors have suggested that the remnant is more
distant (12--14~kpc), on the basis of its optical morphology
and H$\alpha$ to [S {\sc II}] ratio, which are more typical of older
(hence larger) remnants like the Monoceros Ring (\cite{em79}; \cite{kw79}).
More recently, Rho (1995) \nocite{rho95} concluded from
the neutral hydrogen absorption component of the remnant's X-ray spectrum 
(see below)
that the distance is $\sim$7~kpc, consistent with the pulsar distance,
although this method has large uncertainties.  
Independently, Rosado \etal (1997) \nocite{racm97} argued that the remnant 
must be at $\sim$7~kpc, on the basis 
of the kinematics of the optical emission,
although they too concede that the estimate is uncertain given the complexity
of the field.  Thus, overall, the estimated distances to the remnant are
generally consistent with that of the pulsar within the substantial 
uncertainties; an association is therefore plausible.

\begin{figure}
\plotone{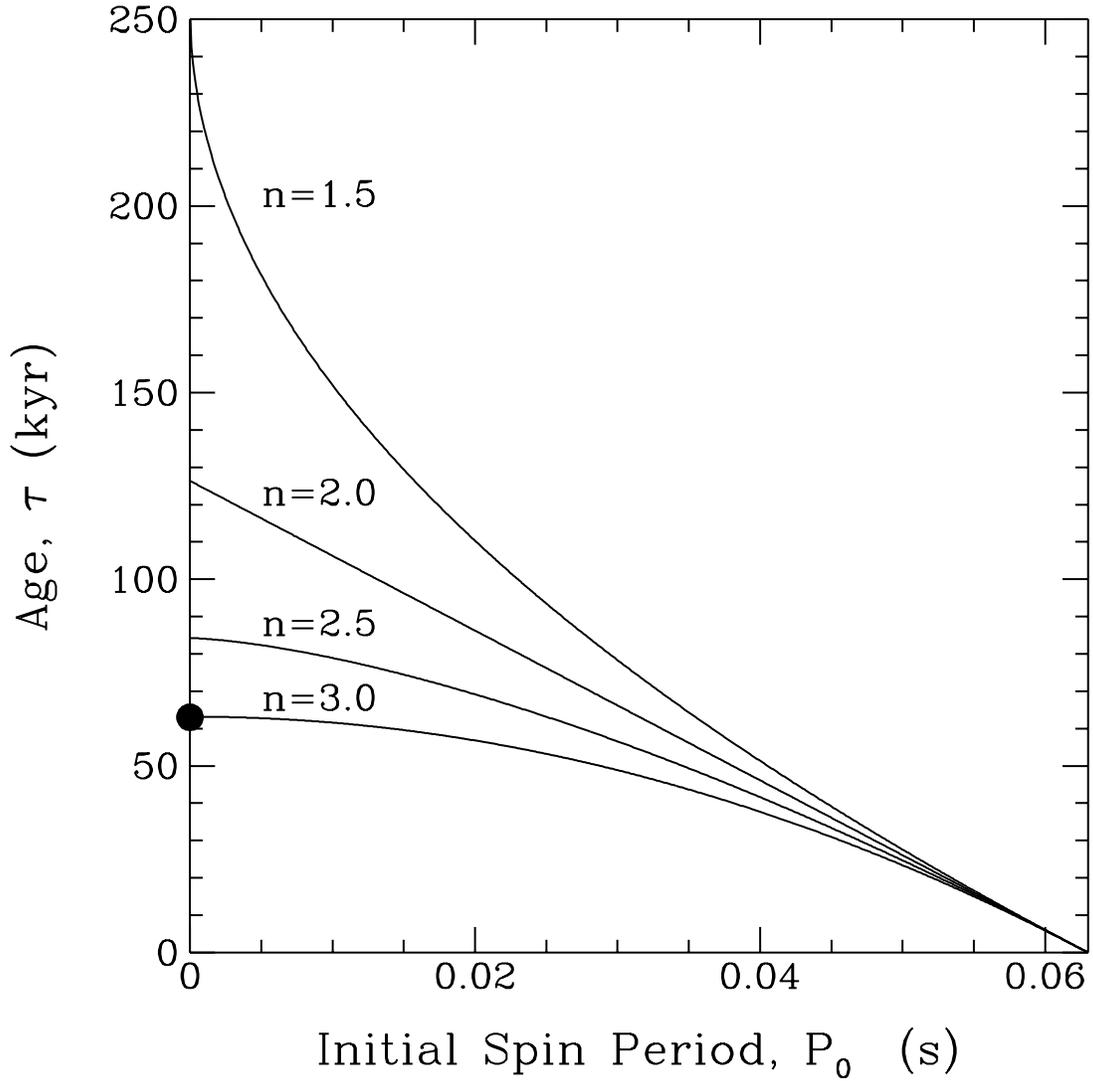}
\caption{The true age $\tau$ of \psr\ as a function of the assumed initial spin period $P_0$
for four values of the braking index $n$.  The filled circle represents
the characteristic age  $\tau_c \equiv P/2\dot{P}$, typically assumed for a pulsar's age.}
\label{fig:age}
\end{figure}

The age $\tau$ of a radio pulsar is given by 
\begin{equation}
\tau = \frac{P}{(n-1)\dot{P}}\left[1-\left(\frac{P_0}{P}\right)^{n-1}\right],
\end{equation}
where $P$ is its current spin period, $P_0$ is its spin period at birth,
and $n$ is its braking index.  The braking index $n$ is defined by
the pulsar spin-down $\dot{\nu} = -K \nu^n$, where $\nu \equiv 1/P$, and
$K$ is a positive constant that depends  on the magnetic dipole moment and
moment of inertia of the rotating neutron star (\cite{mt77}).  It is easy to 
show that $n = \nu \ddot{\nu}/ \dot{\nu}^2$, and hence can be 
determined from timing observations in the absence of strong timing noise.
The characteristic age is defined as $\tau_c \equiv P/2 \dot{P}$, which
assumes that $n=3$ (true for a simple dipole) and $P_0 << P$.  For
\psr, $\tau_c = 63$~kyr.  However, the short spin period for \psr\
compared with other, younger pulsars (e.g. PSR B1509$-$58, $\tau_c =1.5$~kyr,
$P=150$~ms) suggests that $P_0 << P$ does not necessarily hold in this case,
and that the true age may be smaller.  Alternatively, if the braking index
$n < 3$, as is the case for all pulsars for which it has been measured,
then $\tau_c$ is an underestimate.  Notable is the recent
measurement by Lyne \etal (1996b)\nocite{lpgc96} of $n=1.6 \pm 0.4$ for the 
Vela pulsar. 
Figure~\ref{fig:age} shows how the true age of \psr\ depends on $P_0$ for 
four values of $n$.  For small $P_0$, the pulsar's age may be anywhere
between 63 and 250~kyr.  For $P_0 \simeq P$, the pulsar's age
must be smaller than 63~kyr, independent of $n$.  If we assume the pulsar
was born with a spin period of $\sim 20$~ms, as for the Crab pulsar, then
$63 < \tau < 110$~kyr.   

Age estimates for the remnant depend strongly on its distance.  Milne
\etal (1989) \nocite{mck+89} estimated the remnant to be only 2.2~kyr old,
assuming the smallest distance estimate.  Rho (1995) suggests the
remnant is somewhat older, $\sim$10~kyr.  The resemblance of the
optical emission to that of the Monoceros Ring, whose age is $\sim
50$~kyr (\cite{lns86}), suggests a much larger age for \snr.  If it is
associated with \psr, the most likely range of true pulsar ages
requires the remnant to have an age significantly larger than the Milne \etal
estimate, more in line with the more recently suggested hypotheses that 
it is at a greater distance.

\begin{figure}
\plotone{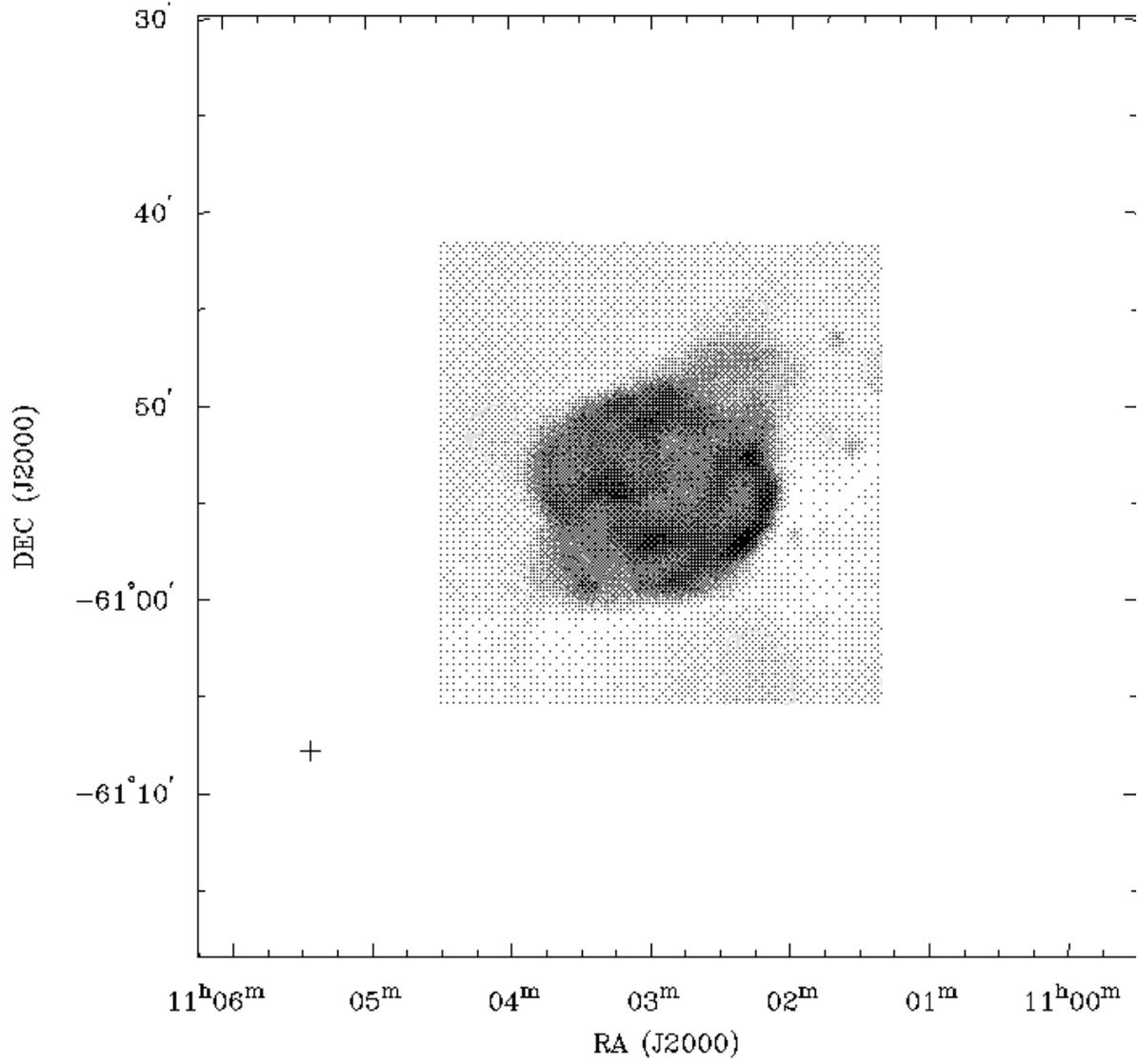}
\caption{MOST radio image of \snr\ at 843~MHz (after \protect\cite{wg96}).
The position of the pulsar is indicated by a cross.}
\label{fig:snr}
\end{figure}

Figure~\ref{fig:snr} shows the location of the pulsar with respect to the
remnant.  It lies just over two remnant radii from the approximate remnant 
geometric center.  Its location, well outside the remnant boundaries, argues 
against an association (\cite{gj95c}).  However, for a distance of 7~kpc, and 
assuming the age of the system to be 63~kyr, the transverse velocity of the 
pulsar, if it is associated with the remnant,
is $\sim 650$~km~s$^{-1}$, larger than the mean pulsar transverse velocity 
(\cite{ll94}), but much less than has been suggested for pulsars
in other proposed associations (e.g. \cite{fk91}; \cite{mkj+91}; \cite{car93}), and 
well within the range of measured pulsar velocities (\cite{ll94}).
Thus, that the pulsar lies well
outside the remnant does not necessarily rule out an association;
it would require the pulsar transverse velocity to exceed the mean
remnant expansion velocity by more than a factor of two, not unreasonable
if the remnant is expanding into a dense environment, as is suggested
by its axisymmetric morphology.
Radio maps of the region closer to the pulsar show no
evidence for any emission that might suggest another, closer supernova 
remnant, or a bow shock nebula (A. Green, personal communication).
Perhaps interestingly, the pulsar's inferred trajectory approximately
bisects the remnant along its line of symmetry.

Seward (1990) \nocite{sew90} and Rho (1995) presented X-ray images 
of \snr\ that show that the emission is
centrally peaked.  This is in contrast to the radio morphology which is
more shell-like.  This suggests that \snr\ is like the supernova remnant W44
(\cite{rpsh94}), which, like W28 and 3C400.2 (\cite{lbmw91}), has
centrally peaked X-ray emission but shell-like radio morphology.
The central X-ray emission in these remnants is thermal.  One possible
interpretation is that the emission is due to the evaporation of dense cloudlets
that survived the initial blast wave, rather than a central neutron
star, as is the case for remnants with centrally peaked radio and non-thermal
X-ray emission, such as the Crab nebula.
Since the morphology suggests that this is also true of \snr, there
is no evidence for a central point source that would argue against
an association with \psr.  We note also that W44, W28, and 3C400.2
are relatively old remnants, all having estimated ages greater than 10~kyr.

A measurement of the proper motion for \psr\ is highly desirable for
determining whether it is associated with \snr.
If the association is real, the pulsar proper motion should be 
$\sim$22~(63~kyr/$\tau$)~mas~yr$^{-1}$, independent of the distance.
A timing proper motion will not be forthcoming, given the large amount 
of timing noise exhibited by the pulsar (Figure~\ref{fig:res}).  Also,
its low flux density (see Table~\ref{ta:parms}) will make interferometric
observations using currently available telescopes difficult, although
pulse gating may improve the feasibility.
A measurement of the pulsar's scintillation speed
through observations of its radio dynamic spectrum, a technique recently
used by Nicastro, Johnston \& Koribalski (1996) \nocite{njk96} to argue 
against an association between the radio pulsar PSR~B1706$-$44 and the 
supernova remnant G343.1$-$2.3, may provide
some evidence against an association if a small speed is found. 
However, this will be difficult again because of the pulsar's low flux density. 
A large pulsar velocity away from the remnant could be confirmed by
the presence of H$\alpha$ emission from a bow shock nebula 
(c.f. \cite{crl93}), although a lack of such emission could be due
to an absence of ambient neutral hydrogen, and would not disprove an
association.

\subsection{Possible association with $\gamma$-ray source 2EG J1103$-$6106}

At a distance of 7~kpc, given its large spin-down
luminosity (Table~\ref{ta:parms}), \psr\ ranks 19th in a
list of rotation-powered pulsars ordered by
$\dot{E}/d^2$.  Six of the seven top spots are held by known 
$\gamma$-ray pulsars (the seventh being the millisecond pulsar
PSR J0437$-$4715), while
most of the top 30 are known X-ray sources.  On this
list, \psr\ ranks 15 spots higher than the known
X-ray and $\gamma$-ray
pulsar PSR~B1055$-$52 (\cite{ch83}; \cite{fbb+93}).  Thus, \psr\
is a good candidate to be an observable high-energy emitter.

In fact, the radio timing position of \psr\ (Table~\ref{ta:parms}) 
lies well inside the 95\%
confidence $49^{\prime} \times 32^{\prime}$ error ellipse of
the second EGRET catalog source \gro\ (\cite{tbd+95}).  This 
$\gamma$-ray source was referred to in the first EGRET catalog as 
GRO~J1110$-$60 (\cite{fbc+94}), and is near, but outside, the error box
of the second COS-B catalog source 2CG 288$-$00
(\cite{sbb+81}).  Reported $E>100$~MeV fluxes of \gro\ show no evidence for
significant variability, consistent with its interpretation as
a rotation-powered pulsar (\cite{tbd+95}; \cite{rbf+95}).  If the sources
are associated, the estimated mean flux of \gro\ suggests that 
\psr\ converts approximately 3\% of its spin-down
luminosity to high-energy $\gamma$-rays for a beaming angle of 1.0~sr,
comparable with the efficiencies of the Vela pulsar and PSR B1706$-$44
(\cite{tab+92}; \cite{ghc88}).

There are 18 radio pulsars within 10$^{\circ}$ of the Galactic
plane that have higher $\dot{E}/d^2$ than PSR B1055$-$52 for which
pulsations have not yet been detected by EGRET, omitting millisecond
pulsars.  With the discovery of \psr, four of these
lie within the 99\% confidence contours of
unidentified EGRET sources.  By contrast, of 268 known radio pulsars
within 10$^{\circ}$ of the Galactic plane whose energetics should be
below the EGRET threshold for detection, only two lie within the 99\%
confidence contours of unidentified EGRET sources (\cite{fie95}).
Assuming that this control group is spatially distributed like the young
pulsars, using Poisson statistics, the probability for
four coincidences among the 18 energetic pulsars is $\sim 1 \times 10^{-5}$.
Even conservatively accounting for the possibility that the control
group is less concentrated near the Galactic plane (for example, by
assigning it a significantly larger mean z-height), we find that the
probability for four coincidences must be under $\sim$1\%, although 
exact probabilities
are difficult to estimate given the uncertainties in pulsar distances and
spatial distributions, and in unidentified EGRET source properties.
Even so, the evidence argues strongly that at least three of the
four coincidences of high $\dot{E}/d^2$ pulsars with the unidentified
EGRET sources are real. 
Furthermore, Yadigaroglu \& Romani (1997) \nocite{yr97} showed that most of the 
unidentified low-latitude EGRET sources such as \gro\ are likely to be 
young pulsars like \psr.  We therefore conclude
that the association between \psr\ and \gro\
is likely.  However, only the detection of $\gamma$-ray pulsations at the radio
period will demonstrate the association unambiguously.

Several authors have argued that the $\gamma$-ray source \gro, as well as 
GRO~J1110$-$60 and 2CG 288$-$00, are associated with
the Carina complex, which includes the peculiar star $\eta$Car, open
clusters Tr 16, Tr 14, and several OB associations, with the
$\gamma$-rays being produced by cosmic ray interactions in the
intercluster gas, or by $\eta$Car itself (\cite{mfb84}; \cite{bf93}; \cite{mpn+96}).
The identification of \gro\ with \psr\ does not necessarily preclude
these interpretations, since \gro\ may be a composite of several
sources.  Indeed there is marginal evidence that its emission is
extended (\cite{sbb+81}; \cite{tbd+95}).  Nevertheless, the discovery of a
luminous young pulsar near the $\gamma$-ray source casts some doubts on
alternative interpretations. Sturner \& Dermer (1995) \nocite{sd95}
suggested that GRO~J1110$-$60 is associated with the supernova remnant
G291.0$-$0.1 (MSH~11$-$6{\it 2}), with the emission a result of cosmic
ray interactions with the remnant.  With the revisions made in the
second EGRET catalog, the source is now closer to the position of
\psr\ and \snr; this and the discovery of \psr\ suggest that their proposed
model is not relevant to this particular $\gamma$-ray source.

Kaaret \& Cottam (1996) \nocite{kc96} suggested that
\gro\ is a young pulsar associated with the OB association Car 2.
The measured distance to the association is 2.2~kpc (\cite{me95}), which
is inconsistent with the DM-derived distance of 7~kpc for \psr, suggesting
that the association lies in the foreground.  If the pulsar actually is in
the cluster, its association with \snr\ is doubtful because the remnant
dimensions would suggest that it is much younger than the pulsar, and it would
be hard to understand the pulsar's position so far outside the remnant
since the latter would have had less time to decelerate.
In this case, the pulsar should be a bright X-ray source, with 
X-ray luminosity $\sim 2 \times 10^{-11}$~erg~s$^{-1}$~cm$^{-2}$
(\cite{sw88}), and should be easily detected by X-ray satellites such as
{\it ASCA}.

\section{Conclusions}

We have reported the discovery and follow-up timing observations of 
\psr\ which show it to
be young and energetic.  We have considered its proximity to the
supernova remnant \snr\ and show that an association between the two
is possible, and could be confirmed or disproved by proper-motion measurements.
We have also considered a 
possible association of \psr\ with the EGRET source \gro\ and conclude
that it is likely.  

It is remarkable that this interesting pulsar was found serendipitously
in a search unrelated to either EGRET sources or supernova remnants,
while recent targeted searches of both have
been done but have met very limited success.
(e.g. \cite{kmj+96}; \cite{gra+96}; \cite{ns96}).  That \psr\ was missed
by a survey including \snr\ is not surprising given the pulsar's low flux 
density and large angular displacement from the remnant.  
If the association between \psr\ and \snr\ is one day proven, it,
and other plausible pulsar/SNR associations in which the pulsar lies 
outside the remnant boundaries (see \cite{kas96} for a review), 
would argue strongly that 
care must be taken to search a large area around the remnant, not just inside.
Either way, the discovery of \psr\ suggests that deeper searches 
of the error boxes of unidentified EGRET sources for radio pulsars are 
warranted.  

\acknowledgements

We thank Joseph Fierro for helpful discussions regarding EGRET source
coincidences with radio pulsars, Shri Kulkarni for his role in
building the Caltech correlator, Anne Green for useful discussions,
and Bryan Gaensler for help with the remnant image.
The filter-bank systems used in the observations were constructed at
the University of Manchester, Jodrell Bank.  Part of this research was
carried out at the Jet Propulsion Laboratory, California Institute of
Technology, under contract with the National Aeronautics and Space
Administration.  VMK received support from NRAO, and from Hubble
Fellowship grant number HF-1061.01-94A from the Space Telescope Science
Institute, which is operated by the Association of Universities for
Research in Astronomy, Inc., under NASA contract NAS5-26555.

\newpage


\newpage

\clearpage
\begin{deluxetable}{ll}
\tablecaption{Astrometric, Spin, and Radio Parameters for \psr.}
\tablehead{\colhead{Parameter                } & \colhead{Value}}
\startdata
Right Ascension, $\alpha$ (J2000) & 11$^{\rm h}$ 05$^{\rm m}$ 26$^{\rm s}$.07(7) \nl
Declination, $\delta$ (J2000) & $-$61$^{\circ}$ ~07$^{\prime}$ ~52$^{\prime \prime}$.1(4) \nl
Galactic Latitude, $l$ & 290$^{\circ}$.4896(2) \nl
Galactic Longitude, $b$ & $-$0$^{\circ}$.8465(1) \nl
Period, $P$ & 0.063191252792(3) s \nl
Period Derivative, $\dot{P}$ & 15.80466(12) $\times$ 10$^{-15}$ \nl
Dispersion Measure, DM & 271.01(2) pc~cm$^{-3}$\nl
Epoch of Period & MJD 49545.0000 \nl
R.M.S. timing residual & 6.2 ms\nl
Surface Magnetic Field Strength, $B$ & $1.0 \times 10^{12}$ G \nl
Characteristic Age, $\tau_c$ & 63,350 yr \nl
Spin-Down Luminosity, $\dot{E}$ & $2.5 \times 10^{36}$ erg~s$^{-1}$ \nl
Flux Density at 660~MHz\tablenotemark{1} & 4.1(7) mJy \nl
Flux Density at 1420~MHz\tablenotemark{2} & 1.84(14) mJy \nl
Flux Density at 1650~MHz\tablenotemark{3} & 1.58(6) mJy \nl 
Spectral Index &  $-$1.36(1) \nl
50\% width at 1650 MHz, W50 & 3.4 ms = 53.9 mP \nl
10\% width at 1650 MHz, W10 & 4.8 ms = 76.0 mP \nl
\tablecomments{(1) reported flux is the mean of those measured at 3 epochs \\ 
(2) reported flux is the mean of that measured at 24 epochs \\
(3) reported flux is the mean of that measured at 16 epochs}
\enddata
\label{ta:parms}
\end{deluxetable}

\end{document}